\documentstyle[multicol,aps,epsf]{revtex}
\newcommand{\beq}{\begin{equation}}
\newcommand{\eeq}{\end{equation}}
\begin{document}
\draft
\baselineskip=16pt

\title{Theory of high-T$_{\bf c}$ superconductivity based
on the fermion-condensation quantum phase transition}

\author{M.Ya. Amusia$^{a,b}$, S.A. Artamonov$^{c}$, and V.R.
Shaginyan$^{c}$ \footnote{E--mail: vrshag@thd.pnpi.spb.ru}}
\address{$^{a\,}$The
Racah Institute of Physics, the Hebrew University, Jerusalem 91904,
Israel;\\ $^{b\,}$Physical-Technical Institute, Russian Academy of
Sciences, 194021 St. Petersburg, Russia;\\ $^{c\,}$ Petersburg Nuclear
Physics Institute, Russian Academy of Sciences, Gatchina, 188350,
Russia}
\maketitle

\begin{abstract}
A theory of high temperature superconductivity based on the
combination of the fermion-condensation quantum phase transition and
the conventional theory of superconductivity is presented. This theory
describes maximum values of the superconducting gap which
can be as big as $\Delta_1\sim 0.1\varepsilon_F$, with $\varepsilon_F$
being the Fermi level. We show that the critical
temperature $2T_c\simeq\Delta_1$. If there exists the pseudogap above
$T_c$ then $2T^*\simeq\Delta_1$, and $T^*$ is the temperature at
which the pseudogap vanishes. A discontinuity in the specific heat at
$T_c$ is calculated. The transition from conventional superconductors
to high-$T_c$ ones as a function of the doping level is investigated.
The single-particle excitations and their lineshape are also
considered.  \end{abstract}


The explanation of the large values of the critical
temperature $T_c$, of the maximum value of the superconducting gap
$\Delta_1$ and of the relation between $\Delta_1$ and the temperature
$T^*$ at which the pseudogap vanishes are, as years before, among the
main problems in the physics of high-temperature superconductivity. To
solve them, one needs to know the single-particle spectra of
corresponding metals. Recent studies of photoemission spectra
discovered an energy scale in the spectrum of low-energy electrons in
copper oxides, which manifests itself as a kink in the
single-particle spectra \cite{vall,blk,krc,lanz}. As a result, the
spectra in the energy range (-200---0) meV can be described by two
straight lines intersecting at the binding energy
$E_0\sim(50-70)$ meV \cite{blk,krc}.
The existence of the energy scale $E_0$ could be attributed
to the interaction between electrons and the collective
excitations, for instance, phonons \cite{lanz}. On the other hand,
the analysis of the experimental data on the single-particle electron
spectra demonstrates that the perturbation of the spectra by phonons or
other collective states is in fact very small, therefore, the
corresponding state of electrons has to be described as a strongly
collectivized quantum state and was named ``quantum protectorate''
\cite{rlp,pa}. Thus, the interpretation of the above mentioned kink
as a consequence of electron-phonon interaction can very likely
be in contradiction with the quantum protectorate concept. To describe
the single-particle spectra and the kink, the assumption can be used
that the electron system of high-$T_c$ superconductor has undergone the
fermion-condensation quantum phase transition (FCQPT). This transition
serves as a boundary separating the normal Fermi liquid from the
strongly correlated liquid of a new type \cite{ks,vol} and fulfills the
quantum protectorate requirements \cite{ms}. The FCQPT appears in
many-electron systems at relatively low density, when the effective
interaction constant becomes sufficiently large. In ordinary electron
liquid, this constant is directly proportional to the dimensionless
parameter $r_s\sim 1/p_Fa_B$, where $a_B$ is the Bohr radius and $p_F$
is the Fermi momentum. The FCQPT appears at a certain value $r_s$,
$r_s=r_{FC}$, and precedes formation of charge-density waves or
stripes \cite{ksz}, which are observed in underdoped samples of copper
oxides \cite{grun}. This is why the formation of the
FCQPT in copper oxides can be considered as a quite determinate process
stemming from general properties of a low-density electron liquid
\cite{ms}.

In this letter we address the mentioned above problems in
the physics of high-temperature superconductivity and demonstrate
that these problems can be resolved in a theory based on the
combination of the FCQPT and the conventional theory of
superconductivity. We show that the FCQPT manifests itself in large
values of $\Delta_1$, $T_c$ and $T^*$.  We trace also the transition
from conventional superconductors to high-$T_c$ ones as a function of
the parameter $r_s$, or as a function of the doping level.

At $T=0$, the ground state energy
$E_{gs}[\kappa({\bf p}),n({\bf p})]$ of two-dimensional electron
liquid is a functional of the order parameter of the superconducting
state $\kappa({\bf p})$ and of the occupation numbers $n({\bf p})$ and
is determined by the known equation of the weak-coupling theory of
superconductivity \beq E_{gs}=E[n({\bf p})]
+\int \lambda_0V({\bf p}_1,{\bf
p}_2)\kappa({\bf p}_1) \kappa^*({\bf p}_2) \frac{d{\bf p}_1d{\bf
p}_2}{(2\pi)^4}.\eeq Here  $E[n({\bf p})]$ is the ground-state energy
of normal Fermi liquid, $n({\bf p})=v^2({\bf
p})$ and $\kappa({\bf p})=v({\bf p})\sqrt{1-v^2({\bf p})}$.
It is assumed that the pairing interaction $\lambda_0V({\bf
p}_1,{\bf p}_2)$ is weak. Minimizing $E_{gs}$ with
respect to $\kappa({\bf p})$
we obtain the equation connecting the single-particle energy
$\varepsilon({\bf p})$ to $\Delta({\bf p})$,
\beq \varepsilon({\bf p})-\mu=\Delta({\bf p})
\frac{1-2v^2({\bf p})} {2\kappa({\bf p})}.\eeq
The single-particle energy $\varepsilon({\bf p})$ is
determined by the Landau equation,
$\varepsilon({\bf p})=\delta
E[n({\bf p})]/\delta n({\bf p})$ \cite{lan}, and
$\mu$ is chemical potential.
The equation for
superconducting gap $\Delta({\bf p})$
takes form
\beq \Delta({\bf p})
=-\int\lambda_0V({\bf p},{\bf p}_1)\kappa({\bf p}_1)
\frac{d{\bf p}_1}{4\pi^2}
=-\frac{1}{2}\int\lambda_0
V({\bf p},{\bf p}_1) \frac{\Delta({\bf p}_1)}
{\sqrt{(\varepsilon({\bf p}_1)-\mu)^2+\Delta^2({\bf p}_1)}}
\frac{d{\bf p}_1}{4\pi^2}.\eeq
If $\lambda_0\to 0$, then, the gap $\Delta({\bf p})\to
0$, and Eq. (2) reduces to the equation proposed in
\cite{ks} \beq \varepsilon({\bf
p})-\mu=0,\: {\mathrm {if}}\,\,\, 0<n({\bf p})<1;\: p_i\leq p\leq
p_f.\eeq
At $T=0$, Eq. (4) defines a particular state of Fermi liquid
with the fermion condensate (FC) for which the modulus of the order
parameter $|\kappa({\bf p})|$ has finite values in the $L_{FC}$
range of momenta $p_i\leq p\leq p_f$, and $\Delta_1\to 0$ in the
$L_{FC}$. Such a state can be considered as superconducting,
with infinitely small value of $\Delta_1$ so that the
entropy of this state is equal to zero. It is obvious, that
this state, being driven by the quantum
phase transition, disappears at $T>0$ \cite{ms}. When $p_i\to
p_F\to p_f$, Eq. (4) determines the point $r_{FC}$ at which the FCQPT
takes place. It follows from Eq. (4) that the system brakes into
two quasiparticle subsystems: the first subsystem in the $L_{FC}$ range
is occupied by the quasiparticles with the effective mass $M^*_{FC}\to
\infty$, while the second one is occupied by quasiparticles with finite
mass $M^*_L$ and momenta $p<p_i$. If $\lambda_0\neq0$, $\Delta_1$
becomes finite, leading to finite value of the effective mass
$M^*_{FC}$ in $L_{FC}$, which can be obtained from Eq. (2) \cite{ms}
\beq M^*_{FC} \simeq
p_F\frac{p_f-p_i}{2\Delta_1}.\eeq As to the energy scale, it is
determined by the parameter $E_0$:
\beq E_0=\varepsilon({\bf p}_f)-\varepsilon({\bf p}_i)
\simeq2\frac{(p_f-p_F)p_F}{M^*_{FC}}\simeq 2\Delta_1.\eeq Thus, a
system with the FC has the single-particle spectrum  of a universal
form and possesses quantum protectorate features at $T\ll T_f$, with
$T_f$ being a temperature, at which the effect of the FCQPT disappears.

We assume that the range $L_{FC}$ is small, $(p_f-p_F)/p_F\ll1$, and
$2\Delta_1\ll T_f$ so that the order parameter $\kappa({\bf p})$ is
governed mainly by the FC \cite{ms}. To solve Eq.  (2) analytically, we
take the Bardeen-Cooper-Schrieffer (BCS) approximation for the
interaction \cite{bcs}:  $\lambda_0V({\bf p},{\bf p}_1)=-\lambda_0$ if
$|\varepsilon({\bf p})-\mu|\leq \omega_D$, the interaction is
zero outside this region, with $\omega_D$ being the characteristic
phonon energy.  As a result, the gap becomes dependent only on the
temperature, $\Delta({\bf p})=\Delta_1(T)$, being independent of the
momentum, and Eq. (2) takes the form \beq
1=N_{FC}\lambda_0\int_0^{E_0/2}\frac{d\xi} {\sqrt{\xi^2+\Delta_1(0)^2}}
+N_{L}\lambda_0\int_{E_0/2}^{\omega_D}\frac{d\xi}
{\sqrt{\xi^2+\Delta_1(0)^2}}.\eeq
Here we set $\xi=\varepsilon({\bf p})-\mu$ and introduce the density
of states $N_{FC}$ in the $L_{FC}$, or $E_0$, range. As
it follows from Eq. (5), $N_{FC}=(p_f-p_F)p_F/2\pi\Delta_1(0)$.
The density of states $N_{L}$ in the range
$(\omega_D-E_0/2)$ has the standard form $N_{L}=M^*_{L}/2\pi$.
If the energy scale $E_0\to 0$, Eq. (7) reduces to the BCS equation.
On the other hand, assuming that $E_0\leq2\omega_D$ and omitting the
second integral in the right hand side of Eq. (7), we obtain
\beq\Delta_1(0)=\frac{\lambda_0 p_F(p_f-p_F)}{2\pi}\ln(1+\sqrt{2})=
2\beta\varepsilon_F\frac{p_f-p_F}{p_F}\ln(1+\sqrt{2}),\eeq
where the Fermi energy $\varepsilon_F=p_F^2/2M^*_L$, and
dimensionless coupling constant $\beta=\lambda_0 M^*_L/2\pi$.
Taking the usual values of the dimensionless coupling constant
$\beta\simeq 0.3$, and $(p_f-p_F)/p_F\simeq 0.2$, we
get from Eq. (7) the large value of $\Delta_1(0)\sim 0.1\varepsilon_F$,
while for normal metals one has $\Delta_1(0)\sim 10^{-3}\varepsilon_F$.
Taking into account the omitted integral, we obtain
\beq\Delta_1(0)\simeq 2\beta\varepsilon_F
\frac{p_f-p_F}{p_F}\ln(1+\sqrt{2})\left(1+\beta
\ln\frac{2\omega_D}{E_0}\right).\eeq
It is seen from Eq. (9) that the correction due to the second
integral is small, provided $E_0\simeq2\omega_D$. Below we show
that $2T_c\simeq \Delta_1(0)$, which leads to the conclusion that
there is no isotope effect since $\Delta_1$ is independent of
$\omega_D$. But this effect is restored as $E_0\to 0$.
Assuming $E_0\sim\omega_D$ and $E_0>\omega_D$, we see that Eq. (7) has
no standard solutions $\Delta(p)=\Delta_1(0)$ because
$\omega_D<\varepsilon(p\simeq p_f)-\mu$ and the interaction vanishes at
these momenta. The only way to obtain solutions is to restore the
condition $E_0<\omega_D$. For instance, we can define the momentum
$p_D<p_f$ such that \beq \Delta_1(0)=2\beta\varepsilon_F
\frac{p_D-p_F}{p_F}\ln(1+\sqrt{2})=\omega_D,\eeq
while the other part in the $L_{FC}$ range can be occupied by a
gap $\Delta_2$ of the different sign, $\Delta_1(0)/\Delta_2<0$. It
follows from Eq. (10) that the isotope effect is presented.
A more detailed analysis will be published elsewhere.

At $T\to T_c$, Eqs. (5) and (6) are replaced by
the equation, which is valid also at $T_c\leq T\ll T_f$ \cite{ms}
\beq M^*_{FC}\simeq p_F\frac{p_f-p_i}{4T_c},\,\,\,
E_0\simeq 4T_c;\,\,\,{\mathrm {if}}\,\,\,T_c\leq T:
\,\,\, M^*_{FC}\simeq p_F\frac{p_f-p_i}{4T},\,\,\,
E_0\simeq 4T.\eeq
Equation (7) is replaced by its conventional finite temperature
generalization
\begin{eqnarray}
1 &=& N_{FC}\lambda_0\int_0^{E_0/2}
\frac{d\xi}
{\sqrt{\xi^2+\Delta_1(T)^2}}
\tanh\frac{\sqrt{\xi^2+\Delta_1(T)^2}}{2T}\nonumber\\
& & + N_{L}\lambda_0\int_{E_0/2}^{\omega_D}
\frac{d\xi}
{\sqrt{\xi^2+\Delta_1(T)^2}}
\tanh\frac{\sqrt{\xi^2+\Delta_1(T)^2}}{2T}.\end{eqnarray}
Putting $\Delta_1(T\to T_c)\to 0$, we obtain from Eq. (12)
\beq 2T_c\simeq \Delta_1(0),\eeq
with $\Delta_1(0)$ being given by Eq. (9).
By comparing Eqs. (5), (11) and (13), we see that $M^*_{FC}$ and
$E_0$ are almost temperature independent at $T\leq
T_c$. Now a few remarks are in order. One can define $T_c$ as the
temperature when $\Delta_1(T_c)\equiv 0$. At $T\geq T_c$, Eq. (12) has
only the trivial solution $\Delta_1\equiv 0$. On the other hand, $T_c$
can be defined as a temperature at which the superconductivity
vanishes.  Thus, we deal with two different definitions, which can lead
to different temperatures. It was shown \cite{sh,ars} that in the case
of the d-wave superconductivity, taking place in the presence of the
FC, there exist a nontrivial solutions of Eq. (12) at $T_c\leq T\leq
T^*$ corresponding to the pseudogap state. It happens when the gap
occupies only such a part of the Fermi surface, which shrinks as the
temperature increases. Here $T^*$ defines the temperature at which
$\Delta_1(T^*)\equiv0$ and the pseudogap state vanishes. The
superconductivity is destroyed at $T_c$, and the ratio $2\Delta_1/T_c$
can vary in a wide range and strongly depends upon the material's
properties, as it follows from consideration given in \cite{sh,ars}.
Therefore, provided there exists the pseudogap above $T_c$, then $T_c$
is to be replaced by $T^*$, and Eq.  (13) takes the form \beq
2T^*\simeq \Delta_1(0).\eeq The ratio $2\Delta_1/T_c$ can reach very
high values. For instance, in the case of
Bi$_2$Sr$_2$CaCu$_2$Q$_{6+\delta}$, where the superconductivity and the
pseudogap are considered to be of the common origin, $2\Delta_1/T_c$
is about 28, while the ratio $2\Delta_1/T^*\simeq 4$, which is also
valid for various cuprates \cite{kug}. Thus, Eq. (14) gives good
description of the experimental data. We remark that Eq. (7) gives also
good description of the maximum gap $\Delta_1$ in the case of the
d-wave superconductivity \cite{sh,ars}, because the different regions
with the maximum absolute value of $\Delta_1$ and the maximal density
of states can be considered as disconnected \cite{abr}. Therefore, the
gap in this region is formed by attractive phonon interaction which is
approximately independent of the momenta.  According to the model
proposed in \cite{ms}, the doping level $x$ is related to the
parameter $r_s$ in the following way: $(x_{FC}-x)\sim
(r_s-r_{FC})\sim (p_f-p_i)/p_F$. The value $x_{FC}$ matches
$r_{FC}$ when defining the point at which the FCQPT takes place. We
assume that the dopant concentration $x_{FC}$ corresponds to the
highly overdoped regime at which slight deviations from the normal
Fermi liquid are observed \cite{val1}.  Then, from Eqs.  (8) and (9)
it follows that $\Delta_1$ is directly proportional to $(x_{FC}-x)$.
From Eq. (14) one finds that the function $T^*(x)$ presents a
straight line crossing the abscissa at the point $(x_{FC}\simeq x)$,
while in the vicinity of this point $T^*$ merges with $T_c$ and both
of them tends to zero.

Now we turn to the calculations of the gap and the specific heat at
the temperatures $T\to T_c$. It is worth noting that this
consideration is valid provided $T^*=T_c$, otherwise the considered
below discontinuity is smoothed out over the temperature range
$T^{*}-T_c$. For the sake of simplicity, we calculate the main
contribution to the gap and the specific heat coming from the FC.
The function $\Delta_1(T\to T_c)$ is found from Eq. (12)
upon expanding the right hand side of the first integral in powers
of $\Delta_1$ and omitting the contribution from the
second integral on the right hand side of Eq. (12). This procedure
leads to the following equation
\beq \Delta_1(T)\simeq
3.4T_c\sqrt{1-\frac{T}{T_c}}.\eeq
Thus, the gap in the spectrum of the single-particle excitations has
quite usual behavior. To calculate the specific heat, the
conventional expression for the entropy $S$ \cite{bcs} can be used
\beq S=2\int\left[f({\bf p})\ln f({\bf p})
+(1-f({\bf p}))\ln(1-f({\bf p}))\right]\frac{d{\bf p}}{(2\pi)^2},\eeq
where
\beq f({\bf p})=\frac{1}{1+\exp[
E({\bf p})/T]};\,\,\,
E({\bf p})
=\sqrt{(\varepsilon({\bf p})-\mu)^2+\Delta_1^2(T)}.\eeq
The specific heat $C$ is determined by
\begin{eqnarray}
C=T\frac{dS}{dT}&\simeq&4\frac{N_{FC}}{T^2}\int_0^{E_0}
f(E)(1-f(E))\left[E^2+T\Delta_1(T)
\frac{d\Delta_1(T)}{dT}\right]d\xi\nonumber\\
& &+4\frac{N_{L}}{T^2}\int_{E_0}^{\omega_D}
f(E)(1-f(E))\left[E^2+T\Delta_1(T)
\frac{d\Delta_1(T)}{dT}\right]d\xi.
\end{eqnarray}
When deriving Eq. (18) we again use the variable $\xi$ and the
densities of states $N_{FC}$, $N_{L}$, just as before in connection to
Eq. (7), and use the notation $E=\sqrt{\xi^2+\Delta_1^2(T)}$. Equation
(18) predicts the conventional discontinuity $\delta C$ in the specific
heat $C$ at $T_c$ because of the last term in the square brackets of
Eq.  (18).  Upon using Eq. (15) to calculate this term and omitting the
second integral on the right hand side of Eq. (18), we obtain \beq
\delta C\simeq\frac{3}{2\pi}(p_f-p_i)p_F.\eeq In contrast to the
conventional result when the discontinuity is a linear function of
$T_c$, $\delta C$ is independent of the critical temperature $T_c$
because the density of state varies inversely with $T_c$ as it follows
from Eq. (11). Note, that deriving Eq. (19) we take into account
the main contribution coming from the FC. This contribution
vanishes as soon as $E_0\to0$ and the second integral of Eq.
(18) gives the conventional result.

Consider the lineshape $L(q,\omega)$ of the single-particle
spectrum which is a function of two variables. Measurements
carried out at a fixed binding energy $\omega=\omega_0$, where
$\omega_0$ is the energy of a single-particle excitation, determine
the lineshape $L(q,\omega=\omega_0)$ as a function of the momentum $q$.
We have shown above that $M^*_{FC}$ is finite and constant at
$T\leq T_c$. Therefore, at excitation energies $\omega\leq E_0$ the
system behaves like an ordinary superconducting Fermi liquid with the
effective mass given by Eq. (5) \cite{ms}. At $T_c\leq T$ the low
energy effective mass $M^*_{FC}$ is finite and is given by Eq. (11).
Once again, at the energies $\omega<E_0$, the system behaves as a
Fermi liquid, the single-particle spectrum is well defined, while the
width of single-particle excitations is of the order of $T$
\cite{ms,dkss}. This behavior was observed in experiments on
measuring the lineshape at a fixed energy \cite{vall}.
It is pertinent to note that recent measurements of the lineshape
suggest that quasiparticle excitation even in the $(\pi,0)$ region of
the Brillouin zone of Bi$_2$Sr$_2$CaCu$_2$Q$_{8+\delta}$ (Bi2212) are
much better defined then previously believed from earlier Bi2212 data
\cite{feng}. The lineshape can also be determined as a function
$L(q=q_0,\omega)$ at a fixed $q=q_0$.  At small $\omega$, the lineshape
resembles the one considered above, and $L(q=q_0,\omega)$ has a
characteristic maximum and width. At energies $\omega\geq E_0$,
quasiparticles with the mass $M^*_{L}$ come into play, leading to a
growth of the function $L(q=q_0,\omega)$. As a result, the function
$L(q=q_0,\omega)$ possesses the known peak-dip-hump structure
\cite{dess} directly defined by the existence of the two effective
masses $M^*_{FC}$ and $M^*_L$ \cite{ms}.  To have more quantitative and
analytical insight into the problem we use the Kramers-Kr\"{o}nig
transformation to construct the imaginary part
${\mathrm{Im}}\Sigma({\bf p},\varepsilon)$ of the self-energy
$\Sigma({\bf p},\varepsilon)$ starting with the real one
${\mathrm{Re}}\Sigma({\bf p},\varepsilon)$ which defines the effective
mass \cite{mig} \beq \frac{1}{M^*}=\left(\frac{1}{M}+\frac{1}{p_F}
\frac{\partial{\mathrm{Re}}\Sigma}{\partial p}\right)/
\left(1-\frac{\partial{\mathrm{Re}}\Sigma}{\partial
\varepsilon}\right).\eeq
Here $M$ is the bare mass, while the relevant momenta $p$ and
energies $\varepsilon$ are subjected to the conditions:
$|p-p_F|/p_F\ll 1$, and $\varepsilon/\varepsilon_F\ll 1$.
We take ${\mathrm{Re}}\Sigma({\bf p},\varepsilon)$ in the simplest form
which accounts for the change of the effective mass at
the energy scale $E_0$:
\beq {\mathrm{Re}}\Sigma({\bf p},\varepsilon)=-\varepsilon
\frac{M^*_{FC}}{M}+\left(\varepsilon-\frac{E_0}{2}\right)
\frac{M^*_{FC}-M^*_{L}}{M}\left[\theta(\varepsilon-E_0/2)
+\theta(-\varepsilon-E_0/2)\right].\eeq
Here $\theta(\varepsilon)$ is the step function. Note that in
order to ensure a smooth transition from the single-particle
spectrum characterized by $M^*_{FC}$ to the spectrum defined by
$M^*_{L}$ the step function is to be substituted by some smooth
function. Upon inserting
Eq. (21) into Eq. (20) we can check that inside the interval
$(-E_0/2,E_0/2)$ the effective mass $M^*\simeq M^*_{FC}$, and outside
the interval $M^*\simeq M^*_{L}$. By applying the Kramers-Kr\"{o}nig
transformation to ${\mathrm{Re}}\Sigma({\bf p},\varepsilon)$, we
obtain the imaginary part of the self-energy, \beq
{\mathrm{Im}}\Sigma({\bf p},\varepsilon)\sim
\varepsilon^2\frac{M^*_{FC}}{\varepsilon_F M}+
\frac{M^*_{FC}-M^*_L}{M}\left(
\varepsilon\ln\left|\frac{\varepsilon+E_0/2}
{\varepsilon-E_0/2}\right|+
\frac{E_0}{2}\ln\left|\frac{\varepsilon^2-E^2_0/4}
{E^2_0/4}\right|\right).\eeq
We can see from Eq. (22)
that at $\varepsilon/E_0\ll 1$ the imaginary part is proportional
to $\varepsilon^2$; at $2\varepsilon/E_0\simeq 1$
${\mathrm{Im}}\Sigma\sim \varepsilon$; at $E_0/\varepsilon\ll 1$
the main contribution to the imaginary part is approximately
constant. This is the behavior that gives rise to the known
peak-dip-hump structure. Then, it is seen from Eq. (22) that when
$E_0\to 0$ the second term on the right hand side tends to zero, the
single-particle excitations become  better defined resembling
that of a normal Fermi liquid, and the peak-dip-hump structure
eventually vanishes. On the other hand, the quasiparticle amplitude
$a({\bf p})$ is given by \cite{mig}
\beq \frac{1}{a({\bf p})}=1-\frac{\partial
{\mathrm{Re}}\Sigma({\bf p},\varepsilon)}{\partial\varepsilon}.\eeq
It follows from Eq. (20)
that the quasiparticle amplitude $a({\bf p})$ rises as the effective
mass $M^*_{FC}$ decreases.  Since $M^*_{FC}\sim (p_f-p_i)\sim
(x_{FC}-x)$ \cite{ms}, we are led to a conclusion that
the amplitude $a({\bf p})$ rises as the doping level rises, and the
single-particle excitations become better defined in highly overdoped
samples. It is worth noting that such a behavior was observed
experimentally in so highly overdoped Bi2212 that the gap size is
about 10 meV \cite{val1}.  Such a small size of the gap testifies
that the region occupied by the FC is small since $E_0/2\simeq
\Delta_1$.

In conclusion, we have shown that the theory of high temperature
superconductivity based on the fermion-condensation quantum phase
transition and on the conventional theory of superconductivity
permits to describe high values of $T_c$, $T^*$ and of the maximum
value of the gap, which may be as big as $\Delta_1\sim
0.1\varepsilon_F$.  We have also traced the transition from
conventional superconductors to high-$T_c$ and demonstrated
that in the highly overdoped cuprates the single-particle
excitations become much better defined, resembling that of a normal
Fermi liquid.

This work was supported in
part by the Russian Foundation for Basic Research, project no.
01-02-17189.


\begin{thebibliography}{99}

\bibitem{vall} T. Valla {\it et al}., Science {\bf 285}, 2110 (1999);
T. Valla {\it et al}., Phys. Rev. Lett. {\bf 85}, 828 (2000).

\bibitem{blk} P.V. Bogdanov {\it et al}., Phys. Rev. Lett. {\bf 85},
2581 (2000).

\bibitem{krc} A. Kaminski {\it et al}.,
Phys. Rev. Lett. {\bf 86}, 1070 (2001).

\bibitem{lanz} A. Lanzara {\it et al}.,
Nature {\bf 412}, 510 (2001).

\bibitem{rlp} R.B. Laughlin and D. Pines, Proc. Natl. Acad. Sci. USA
{\bf 97}, 28 (2000).

\bibitem{pa} P.W. Anderson, cond-mat/0007185; cond-mat/0007287.

\bibitem{ks} V.A. Khodel and V.R. Shaginyan,
Pis'ma Zh. \'{E}ksp. Teor. Fiz. {\bf 51}, 488 (1990) [JETP Lett.
{\bf 53}, {\bf 51}, 553 (1990).

\bibitem{vol} G. E. Volovik,
Pis'ma Zh. \'{E}ksp. Teor. Fiz. {\bf 53}, 208 (1991) [JETP Lett. {\bf 53},
222 (1991)].

\bibitem{ms} M.Ya. Amusia and V.R. Shaginyan,
Pis'ma Zh. \'{E}ksp. Teor. Fiz.
{\bf 73}, 268 (2001) [JETP Lett. {\bf 73}, 232 (2001)];
M.Ya. Amusia and V.R. Shaginyan, Phys. Rev. B {\bf 63}, 224507
(2001).

\bibitem{ksz} V.A. Khodel, V.R. Shaginyan, and M.V. Zverev,
Pis'ma Zh. \'{E}ksp. Teor. Fiz. {\bf 65}, 242 (1997)
[JETP Lett. {\bf 65}, 253 (1997)].

\bibitem{grun} G. Gr\"{u}ner, {\it Density Waves in Solids}
(Addison-Wesley, Reading, MA, 1994).

\bibitem{lan} L. D. Landau,
Zh. \'{E}ksp. Teor. Fiz. {\bf 30}, 1058 (1956)
[Sov. Phys. JETP {\bf 3}, 920 (1956)].

\bibitem{bcs} J. Bardeen, L.N. Cooper, and J.R. Schrieffer,
Phys. Rev. {\bf 108}, 1175 (1957).

\bibitem{sh} V.R. Shaginyan,
Pis'ma Zh. \'{E}ksp. Teor. Fiz. {\bf 68}, 491 (1998)
[JETP Lett. {\bf 68}, 527 (1998)].

\bibitem{ars} S.A. Artamonov and V.R. Shaginyan,
Zh. \'{E}ksp. Teor. Fiz. {\bf 119}, 331 (2001)
[JETP {\bf 92}, 287 (2001)].

\bibitem{kug} M. Kugler {\it et al}.,
Phys. Rev. Lett. {\bf 86}, 4911 (2001).

\bibitem{abr} A.A. Abrikosov, Phys. Rev. B {\bf 52},
R15738 (1995); A.A. Abrikosov, cond-mat/9912394.

\bibitem{val1} Z. Yusof {\it et al}., cond-mat/01044367.

\bibitem{dkss} J. Dukelsky {\it et al}., Z. Phys. {\bf 102}, 245 (1997).

\bibitem{feng} D.L. Feng {\it et al}., cond-mat/0107073.

\bibitem{dess} D.S. Dessau {\it et al}.,
Phys. Rev. Lett. {\bf 66}, 2160 (1991).

\bibitem{mig} A.B. Migdal, {\it Theory of Finite Fermi Systems and
Applications to Atomic Nuclei} (Benjamin, Reading, MA, 1977).

\end{thebibliography}
\end{document}